\documentclass[aps,pra,twocolumn,groupedaddress, amsmath,amssymb]{revtex4}

\usepackage{graphicx}
\usepackage{dcolumn}
\usepackage{bm}
\usepackage{color}

\newcommand{\bra}[1]{\langle #1|}
\newcommand{\ket}[1]{|#1\rangle}

\begin{document}


\title{Hydrogen and Helium atoms in strong magnetic fields}


\author{Anand Thirumalai\footnote{Electronic address: anand@phas.ubc.ca.}}
\author{Jeremy S. Heyl\footnote{Canada Research Chair, electronic address: heyl@phas.ubc.ca.}}
\affiliation{University of British Columbia, Vancouver, British Columbia, V6T 1Z1}


\date{\today}

\begin{abstract}
The energy levels of hydrogen and helium atoms in strong magnetic fields are calculated in this study. The current work contains estimates of the binding energies of the first few low-lying states of these systems that are improvements upon previous estimates. The methodology involves computing the eigenvalues and eigenvectors of the generalized two-dimensional Hartree-Fock partial differential equations for these one- and two-electron systems in a self-consistent manner. The method described herein is applicable to calculations of atomic structure in magnetic fields of arbitrary strength as it exploits the natural symmetries of the problem without assumptions of any basis functions for expressing the wave functions of the electrons or the commonly employed adiabatic approximation. The method is found to be readily extendable to systems with more than two electrons.
\end{abstract}

\maketitle

\section{\label{sec:intro}Introduction}

The motivation to study atoms in magnetic fields of strength beyond the perturbative regime was in a large part due to the discovery of such fields being present in white dwarf stars \cite{Kemp1970, Angel1978, Angel1981} and neutron stars \cite{Trumper1977,Trumper1978}. The most commonly observed neutron stars - pulsars, have been observed to have magnetic fields on the order of \begin{math}10^{11}\end{math} - \begin{math}10^{13}\end{math}G \cite{Ruder94}. Magnetars \cite{DT1992}, which are strongly magnetized neutron stars, can have magnetic field strengths well in excess of \begin{math}10^{13}\end{math}G. White dwarf stars on the other hand have somewhat less extreme fields, albeit still high, \begin{math}\sim10^{6}\end{math} - \begin{math}10^{9}\end{math}G \cite{Ruder94}. At such high field strengths a Zeeman-type perturbative treatment of the field \cite{Landau} is not possible. The structure of atoms is considerably altered from the low field case. 

Since the 1970's this problem has been tackled by various researchers
using different methods. Using a basis of functions for expanding the
wave function, the problem of the hydrogen atom in a strong magnetic
field was tackled with either a variational approach \cite{CK1972}, or
by attempting to solve the Schr\"{o}dinger equation directly
\cite{Praddaude1972, SV1978, Friedrich1982, WR1980, RWRH1983,
  RWRH1984, Ivanov1988}. Elsewhere, work has also been done with
regard to the treatment of the hydrogen atom as well as molecules and
chains of hydrogen atoms in intense magnetic fields \cite{Lai1992,
  Lai1993, Lai1996}, with applications to neutron star atmospheres; for
magnetic fields in excess of $10^{12}$G. Additionally, accurate
results were obtained for atoms and molecules \cite{Medin2006I} and
infinite chains of atoms \cite{Medin2006II} using
density-functional-theory, once again in magnetic fields of neutron
stars in excess of $10^{12}$G. Elsewhere, accurate results for the
singlet and triplet states of the hydrogen molecule have also been
obtained \cite{Schmelcher97hyd, Schmelcher2000hyd, Schmelcher2001hyd}
with direct relevance to the atmospheres of white dwarfs and neutron
stars in low to intermediate field strengths ($0 - 10^{10}$G). More
recently, accurate results have also been obtained for the hydrogen
atom in strong magnetic fields \cite{VB2008} using a Lagrange-mesh
method \cite{VB2002}. 

Initial attempts for estimating the energies and
wave functions of different states of the helium atom were based upon
a purely variational approach \cite{Cohen1970, Henry1974, Mueller1975,
  Banerjee1974, CRS1974, GK1975, Glasser1975, Larsen1979, VB1989} or
$Z-$dependent perturbation theory \cite{Gadiyak1982}. However, the most accurate and reliable solutions to the helium atom
in a strong magnetic field, thus far, involved using the Hartree-Fock
(HF) technique \cite{Hartree} and initial work yielded accurate
eigenvalues for the first few low-lying states of the helium atom and
helium-like species \cite{Virtamo1976, Proschel1982,
  Thurner1993}. These treatises employed Landau orbitals \cite{Landau}
to describe the motion of the electron perpendicular to the field and
employed the adiabatic approximation by limiting the electrons to
reside only in the ground Landau state. Ivanov \cite{Ivanov1994} in
1994 obtained similar results for the helium atom in strong magnetic
fields using an unrestricted HF technique. Elsewhere,
Quantum-Monte-Carlo methods were employed successfully for determining
the ground and first few excited states of the helium atom in low to
strong magnetic fields \cite{Jones1996, Jones1997, Jones1999}. In the
treatises described above \cite{Virtamo1976, Proschel1982,
  Thurner1993, Ivanov1994, Jones1996, Jones1997, Jones1999}, usually
an approximation was employed for calculating both the direct and the
exchange interactions between the electrons. Such approximations
generally involved finding appropriate expansions that mimicked the
behavior of the inter-electron terms in the Hamiltonian. Necessarily,
such an approach is limited by the accuracy of the expansions
employed. In addition, this increases the complexity of the
computational problem. 

Heyl \& Hernquist \cite{HH1998} in 1998 described both an analytical as well as a numerical approximation for evaluating the effective inter-electronic potentials. These were computationally inexpensive. They adopted Hermite polynomials to describe motion of the electron parallel to the magnetic field. This method was seen to yield accurate results for intense magnetic fields with strengths larger than $10^{12}$G, for hydrogen and helium. More recently, Mori \& Hailey \cite{MH2002} and Mori \& Ho \cite{MH2007} adopted a perturbative approach to treat the exchange terms and higher Landau states with success for helium and other mid-$Z$ atoms in high magnetic field strengths. Schmelcher and co-workers have over recent years, carried out detailed Hartree-Fock studies of the helium atom in strong and intense magnetic fields. They employed both a special basis of functions for expressing the wave functions of the electrons \cite{Schmelcher2003, Schmelcher2003II, Schmelcher2002, Schmelcher2001, Schmelcher2000, Schmelcher1999, Schmelcher_helium2007}, adopting a full configuration-interaction approach as well as a numerical mesh-method for solving the unrestricted HF equations \cite{Ivanov1997, Ivanov1999, Ivanov2000, Ivanov2001JPhB, Ivanov2001EPJD}. The special meshes were so constructed as to facilitate finite-difference calculations in a two-dimensional domain using carefully selected mesh node points \cite{Ivanov2001}. These studies yielded accurate eigenvalues for the ground and first few low-lying states of hydrogen and helium. These estimates were seen to be more accurate than previous estimates of the same \cite{Cohen1970, Henry1974, Mueller1975, Banerjee1974, CRS1974, GK1975, Glasser1975, Larsen1979, VB1989, Gadiyak1982, Virtamo1976, Proschel1982, Thurner1993, Ivanov1994, Ruder94, Jones1996, Jones1997, Jones1999}. More recently, Wang and Qiao \cite{Wang2008} employed a configuration-interaction method with Hylleraas-Gaussian-type basis functions for obtaining the energies of low-lying singlet configurations of the helium atom. Their work was based upon an extension of the method due to the authors in Refs.~\cite{Schmelcher1999, Schmelcher2000} with findings consistent with the same.

In the literature, the number of investigations of the helium atom and helium-like species in the strong or intermediate magnetic field regime is rather small. The estimates of the energy levels of these species are only reasonably accurate, and the computational expense is rather high. However, for most observable neutron stars and many white dwarf stars, the magnetic field strengths lie in the intermediate field regime \cite{Ruder94}. In order to facilitate a proper understanding of the spectra of neutron stars and white dwarf stars, one must necessarily have more stringent bounds on the energy levels of atoms in the atmospheres of these compact objects in the intermediate regime of magnetic field strengths. This is the aim of the current work. The work described herein extends previous work \cite{HH1998}, yielding accurate results for the eigenvalues and eigenvectors of the first few low-lying states of hydrogen and helium, over a wide range of field strengths in the intermediate field regime. The calculated energy eigenvalues are seen to be improvements upon previous estimates. The procedures described herein do not make any assumptions of basis functions and neither are they restricted to the adiabatic approximation. The direct and exchange interactions of the electrons are determined using a computational method similar to that employed in Ref.~\cite{Laaksonen1983}; not relying upon any approximations. The method is readily extendable to many-electron systems and arbitrary field strengths. The overall method is also computationally straightforward to implement.

\section{\label{sec:h-atom}The Hydrogen Atom}

We shall begin with the Hamiltonian for the hydrogen atom in a magnetic field using cylindrical co-ordinates;

\begin{eqnarray}
\hat{H}=-\frac{\hbar^{2}}{2m_{e}}
\left(
\frac{1}{\rho}\frac{\partial}{\partial\rho}
	\left(\rho\frac{\partial}{\partial\rho}\right)+\frac{1}{\rho^{2}}\frac{\partial^{2}}{\partial\phi^{2}}+\frac{\partial^{2}}{\partial z^{2}}\right)\nonumber\\
	+\mu_{B}B\left(l_{z}+2s_{z}\right)+\frac{e^{2}B^{2}}{8m_{e}}\rho^{2}-\frac{Ze^{2}}{4\pi\epsilon_{0}}\frac{1}{|{\vec{r}}|},
\label{eq:1}
\end{eqnarray}
where \begin{math} m_{e} \end{math} is the mass of the electron and \begin{math}B\end{math} the magnetic field strength; the vector \begin{math}\vec{B}\end{math} is oriented along the positive $z-$axis. The remaining symbols have their usual meanings. It is of course implicitly assumed herein that the nucleus is infinitely massive. Let us assume a certain form for the wave function of the single electron;

\begin{equation}
\Psi= \psi(\rho,z)e^{im\phi}\chi(s).
\label{eq:2}
\end{equation}
It can be seen immediately that such a choice precludes the use of a basis of functions for describing the behavior of the electron both parallel and perpendicular to the magnetic field. Thus, the time-independent Schrodinger equation in units of Bohr radii can be written as

\begin{eqnarray}
\left[-\left(
\frac{1}{\rho}\frac{\partial}{\partial\rho}\left(\rho\frac{\partial}{\partial\rho}\right)+\frac{\partial^{2}}{\partial z^{2}}\right)+\frac{m^{2}}{\rho^{2}}+2\beta(m-1)+\right.\nonumber\\
\left.\beta^{2}\rho^{2}-\frac{2Z}{\sqrt{\rho^{2}+z^{2}}}\right]\psi\left(\rho,z\right)\nonumber\\
=\epsilon\psi\left(\rho,z\right),
\label{eq:3}
\end{eqnarray}
where \begin{math}m\end{math} is the azimuthal quantum number. In defining Eq.~(\ref{eq:3}) it has been assumed that the electron spin is anti-aligned with the magnetic field. For the hydrogen atom, a consideration of energetics indicates that the spin-up state is less bound than the corresponding spin-down state \cite{Ruder94}. The eigenenergy of the former can be obtained by adding $4\beta$ (in units of Rydbergs) to the eigenenergy of the latter. Additionally, the energy parameter \begin{math}\epsilon\end{math}, is defined as 
\begin{equation}
\epsilon=\frac{2E}{\alpha^{2}m_{e}c^{2}}=\frac{E}{E_{\infty}},
\label{eq:4}
\end{equation}
where \begin{math}E_{\infty}\end{math} is the Rydberg energy. The
quantity \begin{math}\alpha=e^2/(4\pi\epsilon_{0}\hbar c)\approx1/137\end{math} is the fine structure constant. The parameter \begin{math}\beta\end{math} is defined in the usual way as
\begin{math}
\beta = B/B_0
\end{math},
where \begin{math}B_{0}\end{math} is the critical field strength at which point the transition to the intense magnetic field regime occurs \cite{Ruder94}. This is defined as  $B_{0}=(2\alpha^{2}{m_{e}}^{2}c^{2})/(e\hbar)$. It is to be mentioned that in the current paper, we shall be using SI units for all the physical constants and the magnetic field is taken to be in units of Tesla, or explicitly, $kg/ C \cdot s$, where $C$, represents the units for charge - Coulombs.


Thus, beyond a value of \begin{math}\beta\approx1\end{math} the transition to the intense magnetic field regime occurs and the interaction of the electron with the nucleus becomes progressively less dominant as \begin{math}\beta\end{math} increases. In deriving Eq.~\ref{eq:3}, the definition of the Bohr radius was taken to be given by the expression \begin{math}a_{B}=\hbar/\alpha m_{e}c\end{math}. Based upon the above definition of \begin{math} \beta \end{math}, it is convenient to classify the field strength \cite{Jones96} as low (\begin{math}  \beta \leq 10^{-3} \end{math}), intermediate, also called strong (\begin{math} 10^{-3} \leq \beta \leq 1 \end{math}) and intense or high (\begin{math}  1 \leq \beta \leq \infty \end{math}).

It can be seen that Eq.~(\ref{eq:3}) is a linear second order partial differential equation. In the present study it was solved numerically on a computer using finite-element techniques. For details on the numerical treatment of Eq.~(\ref{eq:3}) see \S\ref{sec:numer}.

\section{\label{sec:he-atom}The Helium Atom}

For calculating the atomic structure of two-electron systems we adopt an iterative method, the so-called \textit{self-consistent field method} \cite{Hartree} which essentially solves the Hartree-Fock equations for the electrons. A short derivation of the key equations are given below assuming a single-configuration form for the atom as described above. 

It is to be noted that no restrictions are imposed on the electrons such as the commonly employed \textit{adiabatic approximation} \cite{Ruder94}. 

\subsection{\label{sec:HF}Derivation of the generalized Hartree-Fock equations in partial differential form}

Let us begin with the Hamiltonian of an $N$-electron atom split into one- and two-body terms;

\begin{equation}
\hat{H}=\sum_{i} h_{i} + \sum_{j\neq i} w\left(r_{i},r_{j}\right).
\label{eq:5}\end{equation}

The first part of the Hamiltonian consisting of one-body interactions is given by the standard prescription (using polar cylindrical co-ordinates);

\begin{eqnarray}
h_{i}=-\frac{\hbar^{2}}{2m_{e}}
\left(
\frac{1}{\rho_{i}}\frac{\partial}{\partial\rho_{i}}
	\left(\rho_{i}\frac{\partial}{\partial\rho_{i}}\right)+\frac{1}{\rho_{i}^{2}}\frac{\partial^{2}}{\partial\phi_{i}^{2}}+\frac{\partial^{2}}{\partial z_{i}^{2}}\right)\nonumber\\
	+\mu_{B}B\left(l_{z}^{i}+2s_{z}^{i}\right)+\frac{e^{2}B^{2}}{8m_{e}}\rho_{i}^{2}-\frac{Ze^{2}}{4\pi\epsilon_{0}}\frac{1}{|{\vec{r_{i}}}|},
\label{eq:6}
\end{eqnarray}
where \begin{math}i=1,2,...,N.\end{math}
The two-body term in the Hamiltonian is simply the Coulomb interaction between the \begin{math}i^{th}\end{math} and \begin{math}j^{th}\end{math} electrons; 
\begin{equation}
w\left(r_{i},r_{j}\right)=\frac{e^{2}}{4\pi\epsilon_{0}}\frac{1}{|\vec{r_{i}}-\vec{r_{j}}|}.
\label{eq:7}
\end{equation}
Let us assume that the wave function of a given configuration of electrons is given by
\begin{equation}
\Psi=A_{N}\left(\psi_{1}, \psi_{2}, \psi_{3}, ..., \psi_{N-1},\psi_{N}\right),
\label{eq:8}
\end{equation}
where \begin{math}A_{N}\end{math} is the anti-symmetrization operator. Thus, it can be seen that a single slater determinant is assumed to represent the atomic configuration of all the electrons. The single particle wave functions are assumed to be of the same form as assumed for the case of the hydrogen atom in Eq. (2); explicitly,
\begin{equation}
\Psi_{i}= \psi_{i}(\rho_{i},z_{i})e^{im\phi_{i}}\chi_{i}(s_{i}),
\label{eq:9}
\end{equation}
where \begin{math}i\end{math} labels each of the $N$ electrons. The single particle wave functions \begin{math}\psi_{i}(\rho_{i},z_{i})\end{math} are taken to be real functions. 

In the current study, we shall look at only the fully spin-polarized
states, with each electron's spin being anti-aligned with the magnetic
field, as these states have an exchange interaction between the
electrons. This exchange between the electrons leads to coupling of
the HF equations at each iteration (see Eq.~(\ref{eq:11})). The
partially spin-polarized states on the other hand, only have the direct
part of the interaction between the electrons, thus the HF equations
become uncoupled at each iteration. The method outlined below concerns itself with the determination of this exchange between the electrons using a computational scheme. Thus, the fully spin-polarized states were considered for the purpose of testing the atomic structure code developed herein. At this juncture, it is to be noted that with decreasing magnetic field strength, it has been found that the fully spin-polarized state is not the local ground state of the helium atom \cite{Ivanov2000}. For example, the authors in Ref.~\cite{Ivanov2000} found that the crossover magnetic field strength for the local ground state of the helium atom corresponded to $\beta_{Z} \approx 0.09$. (The magnetic field strength parameter is defined as \begin{math}\beta_{Z}=B/B_{Z}=\beta/Z^{2}\end{math}. The reference magnetic field strength for nuclear charge \begin{math}Z\end{math}, at which the transition to the intense magnetic field regime occurs, is given by \begin{math}B_{Z}=Z^{2}B_{0}\end{math}, with \begin{math}B_{0}\end{math} as defined earlier). The crossover was found to occur from the fully spin-polarized triplet state $1s_{0}2p_{-1}$, to the partially spin-polarized singlet state $1s^{2}$. Above this value of $ \beta_{Z}$ the triplet state is the ground state. For accurate data of the eigenvalues of this spin-polarized state, as well as for a detailed treatment of ground state crossovers, the reader is referred to Ref.~\cite{Ivanov2000} and references therein. Such a detailed study of crossovers to partially spin-polarized configurations was considered to be beyond the scope of the current paper.  
Presently, writing the generalized Hartree-Fock equations for determining the single particle wave functions \begin{math}\psi_{i}\end{math} we have,
\begin{widetext}
\begin{eqnarray}
h\left(r_{i}\right)\psi_{i}\left(r_{i}\right) + \sum_{j\neq i}\left[\bra{\psi_{j}(r_{j})}w(r_{i},r_{j})\ket{\psi_{j}(r_{j})}\psi_{i}(r_{i})\right.
\left.
-\bra{\psi_{j}(r_{j})}w(r_{i},r_{j})\ket{\psi_{i}(r_{j})}\psi_{j}(r_{i})\right]
=E_{i}\psi_{i}(r_{i}),
\label{eq:10}
\end{eqnarray}
where \begin{math}i =1,2,3, ... , N.\end{math}
Substituting the ansatz given in Eq.~(\ref{eq:9}), the assumed
individual electron wave functions, into Eq.~(\ref{eq:10}) we obtain after rearranging some terms:
\begin{eqnarray}
\left[-\left(
\frac{1}{\rho_{i}}\frac{\partial}{\partial\rho_{i}}\left(\rho_{i}\frac{\partial}{\partial\rho_{i}}\right)+\frac{\partial^{2}}{\partial z_{i}^{2}}\right)
+\frac{m_{i}^{2}}{\rho_{i}^{2}}+2\beta_{Z}(m_{i}-1)
+\beta_{Z}^{2}\rho_{i}^{2}-\frac{2}{\sqrt{\rho_{i}^{2}+z_{i}^{2}}}\right]\psi_{i}\left(\rho_{i},z_{i}\right)e^{im_{i}\phi_{i}}&\nonumber\\
+\frac{2}{Z}\sum_{j\neq i}\left[\bra{\psi_{j}(\rho_{j},z_{j})e^{im_{j}\phi_{j}}}\frac{1}{|\vec{r_{i}}-\vec{r_{j}}|}\ket{\psi_{j}(\rho_{j},z_{j})e^{im_{j}\phi_{j}}}\psi_{i}(\rho_{i},z_{i})e^{im_{i}\phi_{i}}-\right.&\nonumber\\
\left.\bra{\psi_{j}(\rho_{j},z_{j})e^{im_{j}\phi_{j}}}\frac{1}{|\vec{r_{i}}-\vec{r_{j}}|}\ket{\psi_{i}(\rho_{j},z_{j})e^{im_{i}\phi_{j}}}\psi_{j}(\rho_{i},z_{i})e^{im_{j}\phi_{i}}\right]&=\epsilon_{i}\psi_{i}\left(\rho_{i},z_{i}\right)e^{im_{i}\phi_{i}}
\label{eq:11}
\end{eqnarray}
\end{widetext}
where
\begin{math}
i,j=1,2,3,...,N.
\end{math}

It is to be noted that the contribution due to electron spin has been averaged out \textit{a priori}. We have chosen to work in units of Bohr radii along with the definitions given below. Additionally, hereafter the exponential factors with \begin{math} im\phi \end{math} are to be interpreted with the appropriate sign depending upon whether they are written in the bra or in the corresponding ket; a minus sign for the former and a plus sign for the latter. This interpretation should presently be applied to Eq.~\ref{eq:11}.

The Bohr radius for an atom of nuclear charge \begin{math}Z\end{math} is given by \begin{math}a_{B}/Z\end{math}, with \begin{math}a_{B}\end{math} as defined earlier. Finally, the energy parameter is defined as $\epsilon_{i}=E_{i}/E_{Z\infty}$, where $E_{Z\infty}=Z^{2}E_{\infty}$, with \begin{math}E_{\infty}\end{math} as defined in Eq.~(\ref{eq:4}). The above written Eq.~(\ref{eq:11}) represents the $N$-coupled Hartree-Fock equations in partial differential form, for an $N$-electron system with nuclear charge $Z$. The system of equations is solved iteratively; see \S\ref{sec:numer} for numerical details. 

Of key concern in the computation of the eigenvalues and eigenvectors is the determination of the direct and exchange interactions between the electrons. In the current study, these have been dealt with in a manner rather different from earlier treatises. In short, these contributions are essentially extra potentials that add to the existing single particle Hamiltonian, except they are coupled through the wave functions weighting them in a given equation in Eq.~(\ref{eq:11}). The procedure employed for evaluating these potentials is described below.

\subsection{\label{sec:direct}The direct interaction}

The treatment of the direct and exchange potentials considered here and in the subsequent section is based upon rigourously determining these potentials by solving their corresponding elliptic partial differential equations. This is carried out in this study by employing the square of the individual electrons' momentum operators \begin{math} -i \hbar \nabla_{i} \end{math}. The methodology is similar to that developed by the authors in Ref.~\cite{Laaksonen1983} and details of the method employed in the current study are given below.

Let us first examine the integral representing the direct interaction between the electrons; 

\begin{equation}
\Phi_{D}=\bra{\psi_{j}(\rho_{j},z_{j})e^{im_{j}\phi_{j}}}\frac{1}{|\vec{r_{i}}-\vec{r_{j}}|}\ket{\psi_{j}(\rho_{j},z_{j})e^{im_{j}\phi_{j}}}.\label{eq:12}
\end{equation}
Let us act on both sides of Eq.~(\ref{eq:12}) with the operator \begin{math}p_{i}^{2} \equiv - \hbar^{2} \nabla^{2}_{i}\end{math}, we then obtain the expression
\begin{widetext}
\begin{equation}
-\hbar^{2}\nabla^{2}_{i}\Phi_{D}=-\hbar^{2}\bra{\psi_{j}(\rho_{j},z_{j})e^{im_{j}\phi_{j}}}-4\pi\delta^{3}(\vec{r_{i}}-\vec{r_{j}})
\ket{\psi_{j}(\rho_{j},z_{j})e^{im_{j}\phi_{j}}}.
\label{eq:13}
\end{equation}
\end{widetext}
This immediately yields
\begin{equation}
\nabla^{2}_{i}\Phi_{D}=-4\pi|\psi_{j}(\rho_{i},z_{i})|^{2}.
\label{eq:14}
\end{equation}
The RHS of Eq.~(\ref{eq:14}) is the square of the \begin{math}j^{th}\end{math} electron's wave function, evaluated using the co-ordinates of the \begin{math}i^{th}\end{math} electron. Noting that Eq.~(\ref{eq:14}) is simply the Laplace equation, it is observed that it is numerically tractable and solved using appropriate boundary conditions to yield the potential \begin{math}\Phi_{D}\end{math}, which is due to the direct interaction between the \begin{math}i^{th}\end{math} and \begin{math}j^{th}\end{math} electrons; see \S\ref{sec:numer} for details on the numerical methods employed. It is to be noted that in contrast to previous work, the problem in Eq.~(\ref{eq:14}) is somewhat simpler, despite having to solve a partial differential equation, as one does not have to find approximate expressions for the mixing terms arising from the interaction between different electronic states and the Coulomb potential. The reader is referred to Refs.~\cite{Proschel1982, Ruder94} and references therein for the different approximation methods employed for obtaining estimates for Eq.~(\ref{eq:12}). 

We now turn our attention to the other two-body term in Eq.~(\ref{eq:11}), the exchange interaction.

\subsection{\label{sec:xchange}The exchange interaction}

We shall follow the same methodology as in our treatment of the direct interaction term. Let us re-write the term in Eq.~(\ref{eq:11}) that relates to the exchange interaction between the \begin{math}i^{th}\end{math} and \begin{math}j^{th}\end{math} electrons,
\begin{equation}
\Phi_{E}=\bra{\psi_{j}(\rho_{j},z_{j})e^{im_{j}\phi_{j}}}\frac{1}{|\vec{r_{i}}-\vec{r_{j}}|}\ket{\psi_{i}(\rho_{j},z_{j})e^{im_{i}\phi_{j}}}.\label{eq:15}
\end{equation}

Again, as in our previous treatment, let us act on both sides of Eq.~(\ref{eq:15}) with the operator \begin{math}\nabla^{2}_{i}\end{math}, this time dropping the redundant factor of \begin{math}-\hbar^{2}\end{math}, to obtain
\begin{widetext}
\begin{equation}
\nabla^{2}_{i}\Phi_{E}=\bra{\psi_{j}(\rho_{j},z_{j})e^{im_{j}\phi_{j}}}-4\pi\delta^{3}(\vec{r_{i}}-\vec{r_{j}})
\ket{\psi_{i}(\rho_{j},z_{j})e^{im_{i}\phi_{j}}}.
\label{eq:16}
\end{equation}
Upon carrying out the integral we get
\begin{equation}
\nabla^{2}_{i}\Phi_{E}=-4\pi\psi_{j}^{*}(\rho_{i},z_{i})\psi_{i}(\rho_{i},z_{i})e^{i(m_{i}-m_{j})\phi_{i}}.
\label{eq:17}
\end{equation}
It is to be noted that by the definition in Eq.~(\ref{eq:9}), \begin{math}\psi^{*}=\psi\end{math} for the spatial part of the individual electron wave functions. At this stage, let us make the ansatz that
\begin{equation}
\Phi_{E}=\alpha_{E}(\rho_{i},z_{i})e^{i(m_{i}-m_{j})\phi_{i}}.
\label{eq:18}
\end{equation}
Let us act on both sides of Eq.~(\ref{eq:18}) with the Laplacian operator,  \begin{math}\nabla^{2}_{i}\end{math}, to obtain
\begin{equation}
\nabla^{2}_{i}\Phi_{E}=
\left[
\frac{1}{\rho_{i}}\frac{\partial}{\partial\rho_{i}}
	\left(\rho_{i}\frac{\partial}{\partial\rho_{i}}\right)-\frac{(m_{i}-m_{j})^{2}}{\rho_{i}^{2}}+\frac{\partial^{2}}{\partial z_{i}^{2}}\right]
\alpha_{E}(\rho_{i},z_{i})e^{i(m_{i}-m_{j})\phi_{i}}.
\label{eq:19}
\end{equation}
It is then a straight-forward matter upon comparing Eq.~(\ref{eq:17}) with Eq.~(\ref{eq:19}) to immediately see that
\begin{equation}
\left[
\frac{1}{\rho_{i}}\frac{\partial}{\partial\rho_{i}}
	\left(\rho_{i}\frac{\partial}{\partial\rho_{i}}\right)-\frac{(m_{i}-m_{j})^{2}}{\rho_{i}^{2}}+\frac{\partial^{2}}{\partial z_{i}^{2}}\right]\alpha_{E}(\rho_{i},z_{i})
=-4\pi\psi_{j}^{*}(\rho_{i},z_{i})\psi_{i}(\rho_{i},z_{i}).
\label{eq:20}
\end{equation}
The elliptical partial differential equation, Eq.~(\ref{eq:20}) is solved numerically and we thus obtain an estimate for the function \begin{math}\alpha_{E}(\rho_{i},z_{i})\end{math}, for each of the pair-wise interactions among the $N$ electrons. Knowing \begin{math}\alpha_{E}(\rho_{i},z_{i})\end{math}, we can obtain \begin{math}\Phi_{E}\end{math} according to Eq.~\ref{eq:18}. Once \begin{math}\Phi_{E}\end{math} and \begin{math}\Phi_{D}\end{math} have been obtained, we can substitute them into Eq.~(\ref{eq:11}) for the potentials due to the direct and exchange interactions respectively, to obtain
\begin{eqnarray}
\left[-\left(
\frac{1}{\rho_{i}}\frac{\partial}{\partial\rho_{i}}\left(\rho_{i}\frac{\partial}{\partial\rho_{i}}\right)+\frac{\partial^{2}}{\partial z_{i}^{2}}\right)
+\frac{m_{i}^{2}}{\rho_{i}^{2}}+2\beta_{Z}(m_{i}-1)
+\beta_{Z}^{2}\rho_{i}^{2}-\frac{2}{\sqrt{\rho_{i}^{2}+z_{i}^{2}}}\right]
\psi_{i}\left(\rho_{i},z_{i}\right)e^{im_{i}\phi_{i}}\nonumber\\
+\frac{2}{Z}\sum_{j\neq i}\left[\Phi_{D}\psi_{i}(\rho_{i},z_{i})e^{im_{i}\phi_{i}}-\alpha_{E}(\rho_{i},z_{i})e^{i(m_{i}-m_{j})\phi_{i}}\psi_{j}(\rho_{i},z_{i})e^{im_{j}\phi_{i}}\right]\nonumber\\
=\epsilon_{i}\psi_{i}\left(\rho_{i},z_{i}\right)e^{im_{i}\phi_{i}}.
\label{eq:21}
\end{eqnarray}
Taking the inner product with \begin{math}\int{d\phi_{i}e^{-im_{i}\phi_{i}}}\end{math} on both sides of Eq.~(\ref{eq:21}) we obtain; writing in a compact form,
\begin{eqnarray}
\left[-\nabla^{2}_{i}(\rho_{i},z_{i})+\frac{m_{i}^{2}}{\rho_{i}^{2}}+2\beta_{Z}(m_{i}-1)
+\beta_{Z}^{2}\rho_{i}^{2}
-\frac{2}{r_i}\right]\psi_{i}\left(\rho_{i},z_{i}\right)
+\frac{2}{Z}\sum_{j\neq
  i}\left[\Phi_{D}\psi_{i}(\rho_{i},z_{i})-\alpha_{E}\psi_{j}(\rho_{i},z_{i})\right]=\epsilon_{i}\psi_{i}\left(\rho_{i},z_{i}\right),\label{eq:22}
\end{eqnarray}
where \begin{math}i,j=1,2,3,...,N\end{math} and
\begin{math}r_i=\sqrt{\rho_i^2+z_i^2}.\end{math}  The total Hartree-Fock energy of
the state is given by
\begin{equation}
\varepsilon_{total}=\sum_{i}\epsilon_{i}-\frac{1}{2}\frac{2}{Z}\sum_{j\neq i}\left[\bra{\psi_{i}(\rho_{i},z_{i})}\Phi_{D}\ket{\psi_{i}(\rho_{i},z_{i})}-\bra{\psi_{i}(\rho_{i},z_{i})}\alpha_{E}\ket{\psi_{j}(\rho_{i},z_{i})}\right].\label{eq:23}
\end{equation}
\end{widetext}
Eq.~(\ref{eq:22}) is the final form for the generalized Hartree-Fock Equations and for two-electron systems we have two equations, however for arbitrary nuclear charge \begin{math}Z\end{math}. 

This completes our derivation of the Hartree-Fock equations for atoms in magnetic fields of arbitrary strength. The following section delineates the numerical procedures employed in the calculation of the energy eigenvalues and eigenfunctions. Thereafter, results are presented and a discussion follows.

\section{\label{sec:numer} Numerical Details }

The eigenvalue problem for the hydrogen atom in Eq.~(\ref{eq:3}) is solved by discretizing the equation and solving the resultant algebraic eigenvalue problem. The discretization is done using the finite-element method (FEM) \cite{FEMbook}. The generalized eigenvalue problem is then solved using a sparse matrix generalized eigensystem solver \cite{Johnson}. This method was found to yield accurate results for the energy eigenvalues of the first few eigenstates with different azimuthal quantum numbers, \begin{math}m\end{math}. Runs were carried out for different values of the magnetic field strength parameter \begin{math} \beta \end{math}, in the range \begin{math} 10^{-2} \leq \beta \leq 10 \end{math}. For testing additional convergence for every run, we employed six different levels of mesh refinement, ranging from coarse to sufficiently fine mesh. The fine mesh calculations for hydrogen and helium for example, respectively, took between a few hours to a few days of computing time on AMD Opteron$^{\textregistered}$ 844 1.8 GHz processors.

For solutions to the helium atom in strong magnetic fields, an \emph{atomic structure software} was developed as a part of this study for the purpose of calculating the energies of different states of multi-electron atoms. The program takes as its input, the number of electrons in the atom \begin{math}n_{e}\end{math}, the nuclear charge \begin{math}Z\end{math}, and the magnetic field strength parameter \begin{math}\beta \end{math} and then proceeds to compute systematically the eigenvalues and eigenfunctions of the coupled system of equations in Eq.~(\ref{eq:22}), according to the iterative procedures described below in brief.

The Eqs.~\ref{eq:14}, \ref{eq:20} and \ref{eq:22} are solved in a
three step process using the iterative self-consistent Hartree-Fock
method \cite{Hartree}. First, an initial estimate is obtained for the
eigenvectors by solving Eq.~\ref{eq:22} without the contributions due
the interaction between the electrons. The second step involves
obtaining estimates for the potentials due to the direct and exchange
interactions amongst the electrons, vis-a-vis, the elliptic partial
differential Eqs.~\ref{eq:14} and \ref{eq:20} are solved using the estimates for the wave functions obtained in the previous step. These estimates are then used to solve for better estimates of the eigenfunctions along with the relevant eigenvalues in Eq.~\ref{eq:22}. The last two steps are iterated in the order described above to obtain progressively better estimates for the eigenvalues and eigenvectors with each iteration. It was observed during our runs that fast convergence was achieved; within the first few iterations. A convergence criterion was employed wherein the difference between the solutions for two consecutive iterations was tested. Typically, a tolerance on the order of \begin{math}10^{-6}\end{math} was employed. Thereafter, the total energy of the Hartree-Fock state under consideration is reported according to Eq.~\ref{eq:23}. Again, as in the case of the hydrogen atom runs were carried out for different values of \begin{math} \beta \end{math}, however with five different levels of mesh refinement in each case for additional convergence testing. 

Once again, the generalized system of eigenvalue equations were solved using appropriate FEM discretization, yielding accurate results for the eigenvalues and eigenvectors of the first few low-lying states of helium. For details on the finite-element method the reader is referred to Ref.~\cite{FEMbook}. It was found that domain compactification was computationally expensive in terms of memory and computation time for obtaining sufficiently accurate results. On the other hand, limiting the domain size in the two orthogonal directions (parallel and perpendicular to the magnetic field) to several Bohr radii (\begin{math}\approx 20\end{math}) was observed to give accurate results for the range of magnetic field strengths considered in this study. Concordantly, the computational expense was many times less in comparison to the former case. Numerical errors arising from truncating the domain were not significant ({\em i.e.} they were smaller than the discretization errors). 

Additionally, the current work does not include relativistic corrections to the energies. For the magnetic field strengths considered herein, the relativistic corrections to the energies were estimated using the scaling formula of the authors in Ref~\cite{Poszwa2004}. Their results for the hydrogen atom were used for this purpose and the corrections were estimated to be on the order of $10^{-6}$ Rydbergs. This was seen to be smaller than the discretization error. Thus, while relativistic corrections are important, it was not possible to account for them accurately in the current study. 

\section{\label{sec:results} Results \& Discussion }

The results from the calculations carried out for hydrogen and helium atoms in strong magnetic fields  (10$^{-2}$ $\leq$ $\beta$, $\beta_Z$ $\leq$ 10) are presented below for three tightly bound states of each atom in this regime.

\subsection{\label{sec:res-H} The Hydrogen Atom }

We solved the time-independent Schr\"{o}dinger equation given in Eq.~(\ref{eq:3}) for different values of the magnetic field strength parameter for the states with azimuthal quantum numbers, \begin{math} m = 0,-1\end{math} and \begin{math} -2 \end{math}. The binding energies are reported in Rydberg units. (The states with \begin{math} m > 0 \end{math} become unbound with increasing magnetic field strength \cite{Ruder94}.) 

\begin{figure}[h]
\begin{center}
\includegraphics[width=3.5in, scale=0.8]{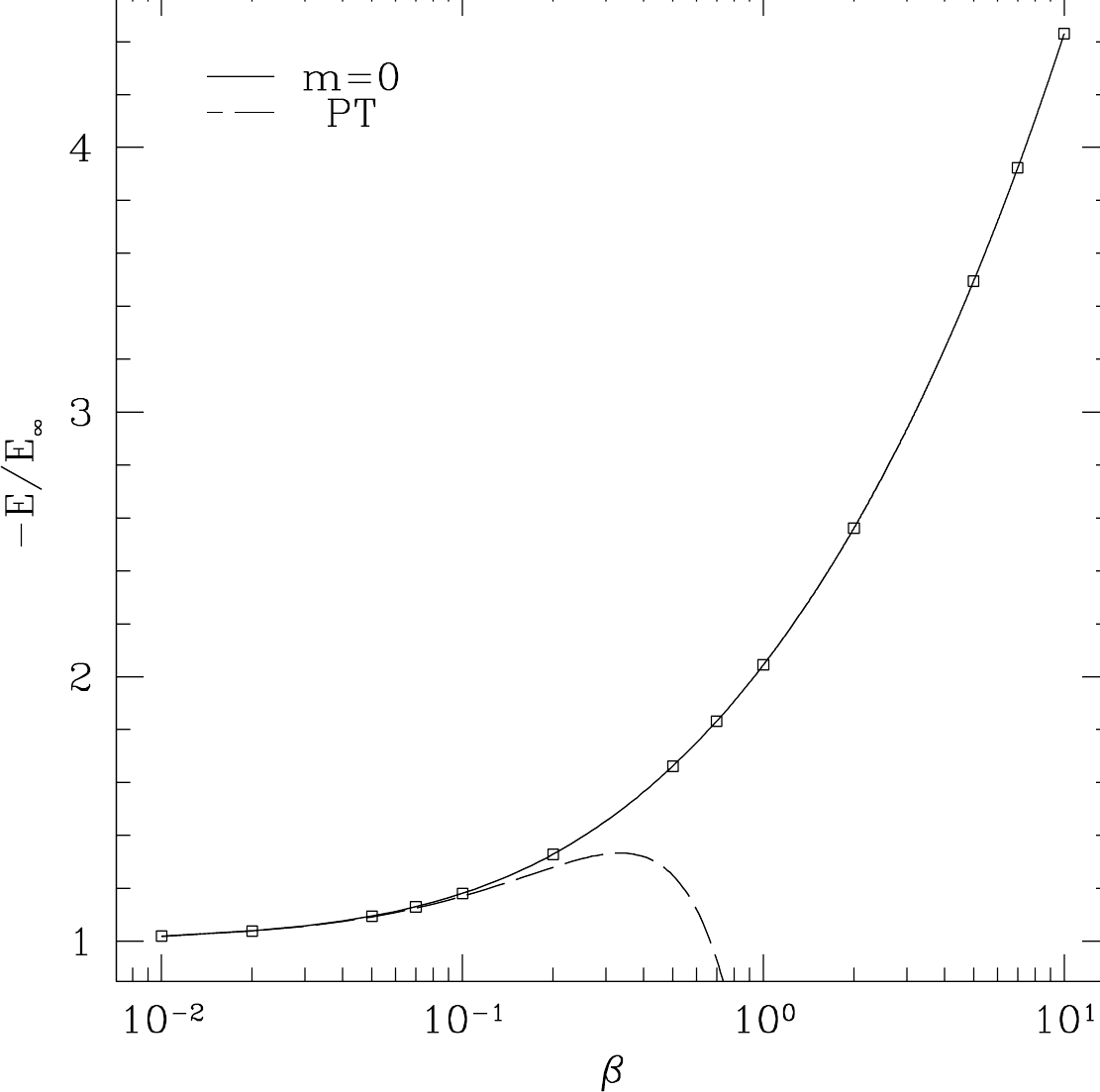}
\end{center}
\caption{Variation in the binding energy of the ground state of hydrogen $m=0$, $\pi=+1$ with the magnetic field strength parameter in the range $10^{-2}\leq\beta\leq 10$. The solid line is the model while the data points are calculated values and the dashed line represents the results from first order perturbation theory. } 
\label{fig:figure1}
\end{figure}

The variation in the binding energy for the state corresponding to \begin{math} m=0 , \pi=+1\end{math} is shown in Figure~\ref{fig:figure1}. The quantity \begin{math} \pi \end{math} indicates parity with respect to the $z-$axis. The data points are eigenvalues obtained from the numerical solution of Eq.~(\ref{eq:3}). The energy eigenvalues are reported as the values corresponding to infinitely fine mesh sizes, or in other words, when the average area of the finite elements approaches zero. A short discussion of this estimation procedure is given later. As can be seen in the figure, the electron becomes progressively more bound as the magnetic field strength increases. The line through the data represents a fit to the data. A rational function was used to model the data in this regime using a robust Levenberg-Marquardt method \cite{NR1992}. 

\begin{table}[t]
  \caption{Coefficients of the different rational
    functions employed for fitting the three states of hydrogen discussed.  The
    maximum fractional error of the eigenvalue relative to the fit from
    $\beta=10^{-4}$ to $\beta=10^3$ follows the
    the list of coefficients.}
\label{tab:Table1}
\begin{tabular}{clcl}
\hline
State & \multicolumn{1}{c}{Coefficients} & State & \multicolumn{1}{c}{Coefficients} \\
\hline
$\begin{array}{c}
1{\rm s}_0 \\
\end{array}$ & 
$\begin{array}{cr}
  a_0=&    3.0849526\\
  a_1=&    6.6571666 \\
  a_2=&  -0.14904847 \\
  a_3=&   0.43176735 \\
  b_0=&    3.0792681 \\
  & 9 \times 10^{-3} 
\end{array}$ &
$\begin{array}{c}
1{\rm s}_0
\end{array}$ & 
$\begin{array}{cr}
a_0 =&   3416640.9     \\
a_1 =&    12197155     \\
a_2 =&   4879198.1     \\
a_3 =&   425959.45     \\
a_4 =&   119957.96     \\
b_0 =&     3418004     \\
b_1 =&   5216732.1     \\
& 4 \times 10^{-4} 
\end{array}$ \\
\hline
$\begin{array}{c}
2{\rm p}_{-1} 
\end{array}$ & 
$\begin{array}{cr}
a_0 =&  0.58201924	 \\
a_1 =&   13.769289	 \\
a_2 =&   17.780336	 \\
a_3 =& -0.52325377	 \\
a_4 =&  0.74135964       \\
b_0 =&   2.3266637	 \\
b_1 =&   18.424245       \\
& 3 \times 10^{-3} 
\end{array}$ & 
$\begin{array}{c}
3{\rm d}_{-2} 
\end{array}$ & 
$\begin{array}{cr}
a_0 =& 0.038324719  \\
a_1 =&   2.7226023  \\
a_2 =&   6.9791245  \\
a_3 =& -0.55680026  \\
a_4 =&   0.4177192  \\
b_0 =&  0.34444067  \\
b_1 =&   6.7184551  \\
& 1 \times 10^{-2} 
\end{array}$  \\
\hline
\end{tabular}
\end{table}

These rational functions accurately model the data in a range of magnetic fields from $\beta=10^{-4}$ to $\beta=10^3$ (twice the range indicated in Figure~\ref{fig:figure1}) and could potentially be used directly in atmosphere models for neutron stars and white dwarf stars. Having such accurate analytical forms for the energies of atoms in strong magnetic fields obviates the need for performing laborious HF calculations, making atmosphere models computationally less intensive. In addition, it becomes possible to analyse spectra of neutron and white dwarf stars with relative ease, at arbitrary field strengths within the intermediate field regime. The rational function fits have the form
\begin{equation}
f(x) = \frac{\sum_{i=0}^n a_i x^i}{x^{n-2} + \sum_{i=0}^{n-3} b_i x^i},
\label{eq:24}
\end{equation} 
where $x=\ln(1+\beta)$. The coefficients and the maximal fitting errors over the entire range $\beta=10^{-4}$ to $\beta=10^3$ are given in Table~\ref{tab:Table1}. The data in the range $\beta=10^{-2}$ to $\beta=10$ are results from calculations of the present study while, to construct the fits for larger and smaller values of $\beta$ than the indicated range, we used the results of Ref.~\cite{Ruder94}. The dashed line in the graph represents a first order perturbation theory calculation. The purpose being to illustrate the fact that perturbation theory breaks down with increasing the magnetic field strength. It is evident upon inspection that the breakdown of perturbation theory occurs within the so called \emph{intermediate} field regime. Thus arises the need for accurate data for the structure of atoms in this regime of magnetic field strengths \begin{math} 10^{-2} \leq \beta \leq 10\end{math}.

\begin{figure}[h]
\begin{center}
\includegraphics[width=3.5in, scale=0.8]{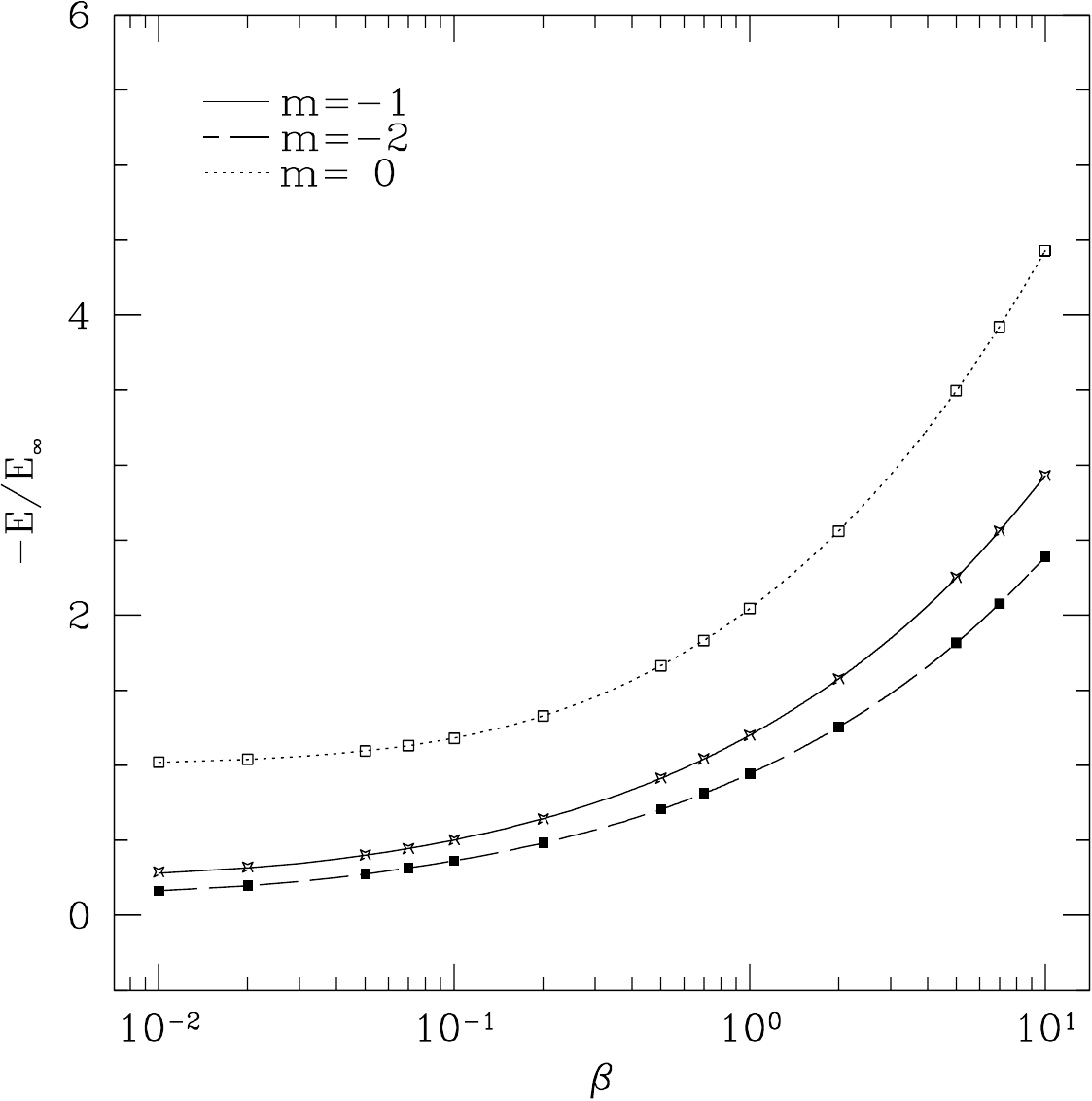}
\end{center}
\caption{Variation in the binding energies of three lowest m- states of hydrogen corresponding to $m=0,-1,-2$, with $\pi=+1$ with the magnetic field strength parameter in the range $10^{-2} \leq \beta \leq 10$.} 
\label{fig:figure2}
\end{figure}

Figure~\ref{fig:figure2} shows variation in the binding energy of the states \begin{math} m=0;  \pi=+1\end{math}, \begin{math} m=-1; \pi=+1\end{math} and \begin{math} m=-2; \pi=+1\end{math}, with increasing magnetic field strength. Again the lines through the data points represent fits to the data. Every data point in these figures was obtained as an estimate corresponding to limit of the finite element size going to zero; see later. It can be seen in Figure~\ref{fig:figure2} that the binding energy of the different states increase dramatically after \begin{math} \beta \approx 1 \end{math}. Thereafter, the binding energy increases at an increasing rate with increasing magnetic field strength.

It is to to be noted that, in all cases, the rational functions were so chosen as to reflect the fact that eventually at large values of the magnetic field strength parameter $\beta$, the energy would be proportional to \begin{math} \ln^2\beta \end{math} \cite{Landau_p460, Ruder94_p77to78, Rau_and_Spruch}. For details of the energy eigenvalues obtained in this study the reader is referred to Table~\ref{tab:Table3} in the appendix to this paper.

Figure~\ref{fig:figure3} shows the dependence of the eigenvalues obtained from the solution of Eq.~(\ref{eq:3}) on the mesh size employed. As mentioned earlier, the calculations were carried out on a finite domain of several Bohr radii in each of the two directions both parallel and perpendicular to the field. While keeping the domain fixed, the number of finite elements constituting the mesh was varied. Runs were performed on each mesh size, for every value of magnetic field strength considered in the study, for each of the three states of hydrogen, $m=0, -1$ and $-2$. For a given value of $\beta$, the eigenvalues were plotted against the average area per finite element in the domain, corresponding to different levels of mesh refinement as shown in the figure. Extrapolation of the data to the limit of zero mesh size, in each case, yielded the eigenvalues that would correspond to an infinitely fine mesh. These values were reported as the calculated data points in the preceding figures. The extrapolation was carried out by employing rational functions; the reader is referred to Ref.~\cite{NR1992} for details regarding the method employed. The average accuracy of the estimate of the asymptotic value was determined to be on the order of \begin{math}3\times10^{-6} \end{math} Rydbergs. These were too small to be shown on the plots. The line through the data points in Figure~\ref{fig:figure3} is merely a guide to the eye.

\begin{figure}[h]
\begin{center}
\includegraphics[width=3.5in, scale=0.8]{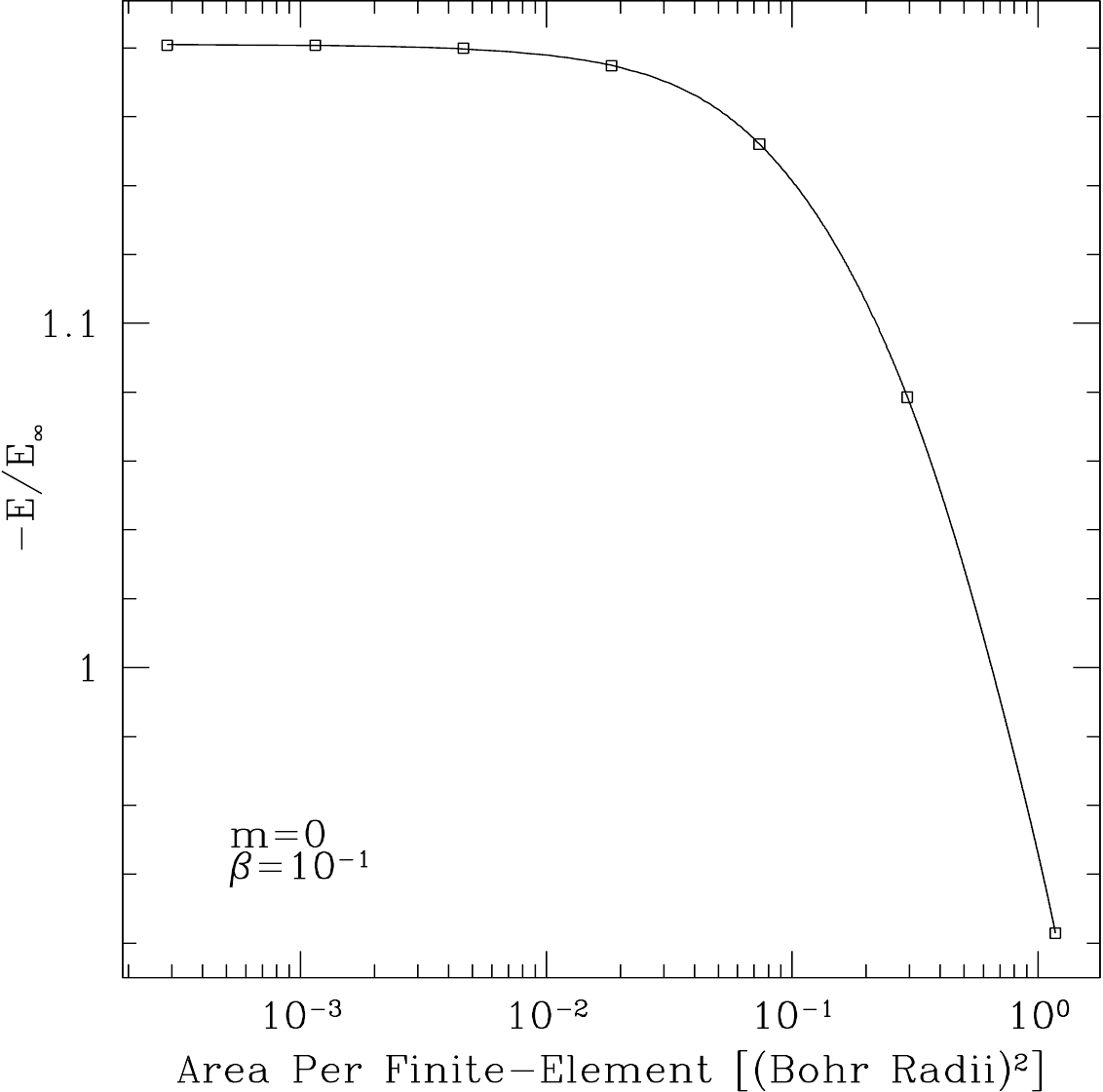}
\end{center}
\caption{Plot of the eigenvalues calculated as a function of the area per finite element in the mesh for the ground state of hydrogen at $\beta$=10$^{-1}$. The eigenvalue approaches an asymptote corresponding to the limit of infinitely fine mesh.}
\label{fig:figure3}
\end{figure}

\begin{figure}[h]
\begin{center}
\includegraphics[width=3.5in, scale=1.0]{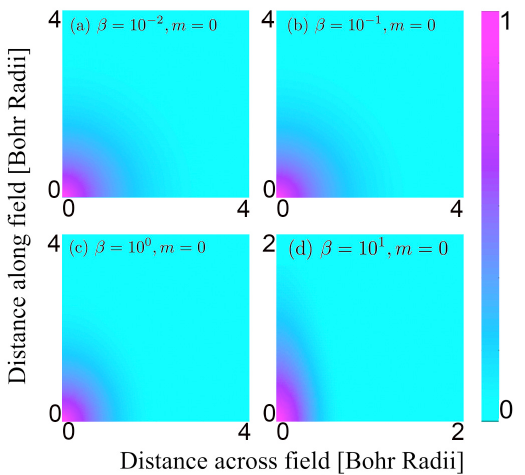}
\end{center}
\caption{Plot of real part of the wave function, i.e. $\psi(\rho,z)$, of the ground state of the hydrogen atom with quantum numbers $m=0$, $s_{z}=-1/2$ and $\pi=+1$ in magnetic fields of strength corresponding to (a) $\beta=10^{-2}$, (b) $\beta=10^{-1}$, (c) $\beta=1$ and (d) $\beta=10$. In figure (d) the relevant portion has been enlarged for better visual inspection.}
\label{fig:figure4}
\end{figure}

The wave functions for the most tightly bound state of
hydrogen \begin{math} m=0, \pi=+1 \end{math} are plotted in
Figure~\ref{fig:figure4} for four different values of the magnetic
field strength, $\beta$. The plots represent a slice through the three
dimensional atom in the upper half of the $r-z-$plane. The complete
representation in three dimensions is the figure of revolution about
the $z-$axis and simultaneously reflected about the $xy-$plane. The
length units are Bohr radii of the hydrogen atom and the horizontal
axis represents the direction perpendicular to the magnetic field
while the vertical axis represents the direction parallel to it,
i.e. the $z-$direction in three dimensional cylindrical polar
co-ordinates. The purpose here is to illustrate the dramatic change
that occurs near $\beta$ $\approx$ 1 when the electron becomes tightly
bound and the binding energy increases dramatically with increasing
$\beta$. It is immediately evident upon inspection that the spherical
symmetry of the atom is clearly broken, as we approach higher magnetic
field strengths and the binding energy increases.

\subsection{\label{sec:res-He} The Helium Atom }

For the helium atom, we solved the set of coupled Hartree-Fock
equations given in Eq.~(\ref{eq:22}), for different values of the magnetic field strength parameter, for the three tightly bound states  \begin{math} M = 0,-1\end{math} and \begin{math} -2 \end{math}. It is to be noted that the binding energies are reported in units of Rydberg energy in the Coulomb potential of nuclear charge \begin{math}Ze\end{math} (i.e., in units of $E_{Z\infty}$, as defined earlier).

At the very outset, as a preliminary test of the Hartree-Fock atomic structure software developed as part of this study, we carried out calculations for the binding energy of the $^{1}S_{0}$ singlet state of neutral helium without any magnetic field ($B=0$ case). However, in this case the wave function of the configuration of electrons is completely symmetrized with respect to the spatial part of the total wave function while the spins of the two electrons are anti-parallel to each other. With these changes, Eq. (24) was solved using the numerical procedures outlined in  \S~\ref{sec:numer} and the energy reported according to Eq. (25). The runs were carried out for different mesh sizes over the domain and the final eigenvalue is reported after extrapolating to the limit of infinitely fine mesh, as described above. The eigenvalue obtained using this procedure was 1.4499 $E_{Z\infty}$, while the most accurately determined result via numerical techniques thus far is 1.4519 $E_{Z\infty}$ \cite{Sims2002}. The difference is about $2 \times 10^{-3}$ $E_{Z\infty}$. This was considered to be sufficiently accurate given the fact that the calculations carried out as part of this study were single-configuration calculations while the result from Ref.~\cite{Sims2002} is essentially a multi-configuration calculation which is a computationally more intensive method. Thus, the atomic structure software developed for this study was considered to be sufficiently accurate for the purposes of the current study. Additionally, we also calculated the HF energies in the limit of zero magnetic field for the three states of helium considered in this study. The energy eigenvalue for the state corresponding to the configuration $1s_02p_{-1}$, with the spins parallel to each other, was computed to be 1.0668 $E_{Z\infty}$, which differs by about $2 \times 10^{-4}$ $E_{Z\infty}$ from accurate calculations including corrections in Ref~\cite{DM98}, wherein they obtained a value of 1.0666 $E_{Z\infty}$. The slight difference in the values is due to discretization errors arising from the finite element method employed in the current study. We also obtained an estimate for the HF energy of the configuration $1s_02p_{0}$, with the spins parallel to each other, to be 1.0641 $E_{Z\infty}$. The difference from the value obtained by researchers in Ref.~\cite{DM98} is about $2.5 \times 10^{-3}$~$E_{Z\infty}$. For the third configuration ($1s_03d_{-2}$, with parallel spins) considered in this study, we obtained the energy eigenvalue of the configuration to be 1.0179 $E_{Z\infty}$. This is less than the value obtained by the researchers in Ref.~\cite{DM98} by about $9.8 \times 10^{-3}$~$E_{Z\infty}$. Since the $d-$orbital is more spread out than the corresponding p-orbital counterparts, it was necessary to sample a greater domain size ($\approx 70 a_{B}$) in the limit of zero magnetic field strength and as a result, due to limitations of computer memory, it was not possible to carry out the calculations at the finest level of mesh refinement. Consequently, the difference between the calculated eigenvalue of the present study and that of Ref.~\cite{DM98} is larger.
 
\begin{figure}[h]
\begin{center}
\includegraphics[width=3.5in, scale=0.8]{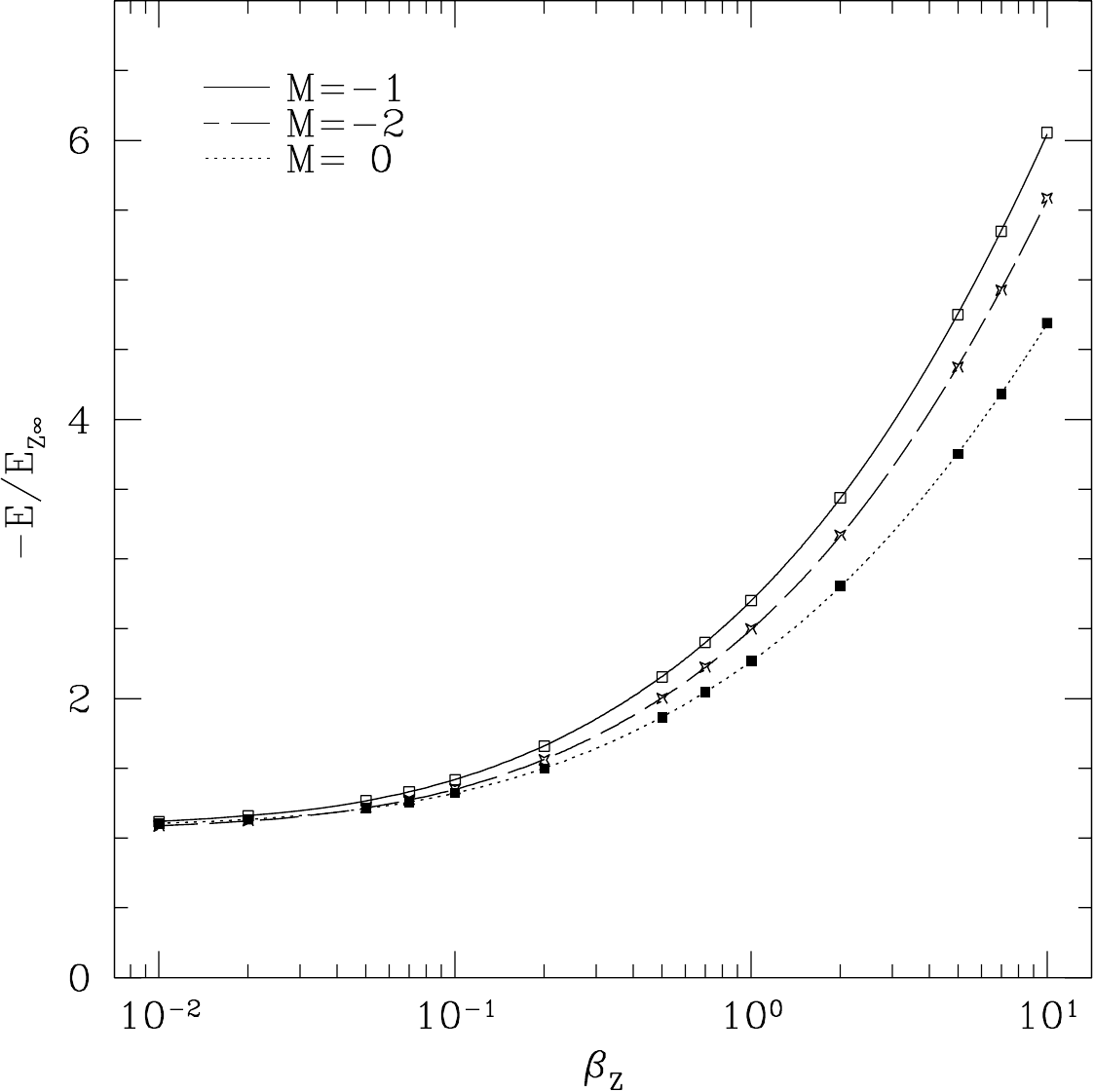}
\end{center}
\caption{Variation in the binding energy of the three tightly bound states of helium $M=-1,-2$ and $0$, with the magnetic field strength parameter in the range $10^{-2}\leq\beta_{Z}\leq10$.} 
\label{fig:figure5}
\end{figure}

The variations in the binding energies for the states corresponding
to \begin{math} M=-1,-2 \end{math} and \begin{math} 0 \end{math}
with \begin{math} S_{z}=-1 \end{math}, are shown in
Figure~\ref{fig:figure5}. The data points are eigenvalues obtained
from the numerical solution of Eq.~(\ref{eq:22}), according to the
numerical procedures described in \S~\ref{sec:numer}. The energy
eigenvalues are once again reported as the values corresponding to an
infinitely fine mesh, estimated according to the discussion above; see
Figure~\ref{fig:figure3} and discussion thereof. The average accuracy
of the estimate of the asymptotic value was determined to be on the
order of \begin{math}2\times10^{-5}\end{math} Rydbergs. Again, for
details regarding the extrapolation method, the reader is referred to
Ref.~\cite{NR1992}. As can be seen in Figure~\ref{fig:figure5}, the
binding energies of the states increase monotonically with increasing
magnetic field strength, \begin{math}\beta_{Z}\end{math}. The lines
through the data represent fits to the data. In addition, a good
measure of the discretisation error is the
difference between the computed eigenvalues for the most finely
refined mesh size employed and the extrapolated result for the mesh
size tending to zero. This error is reported in the number appearing
in the parentheses in Tables~\ref{tab:Table3} and \ref{tab:Table4} in
the appendix to this paper; the number therein corresponds to the
absolute error in the fifth decimal place. Concordantly, the error
bars were too small to be shown on the plots.

Again, a rational function was used to model the data in this regime using a robust Levenberg-Marquardt method \cite{NR1992}. The coefficients of the interpolating functions are given in Table~\ref{tab:Table2}. Again, the rational functions were so chosen as to reflect the fact that eventually, for large values of the magnetic field strength parameter \begin{math} \beta_{Z} \end{math}, the binding energies are proportional to \begin{math} \ln^2 {\beta_{Z}} \end{math}. With this in mind, the aim was to find an accurate analytical function that described the variation in binding energy over a wide range of  \begin{math} \beta_{Z} \end{math}, such that it employed the least number of parameters. Inspection of Table~\ref{tab:Table2} reveals that this was accomplished with five free parameters, for modelling the data in the range of magnetic field strengths $10^{-2} \leq \beta_{Z} \leq 10$. Potentially, as was noted earlier, these analytical functions can be employed with relative ease in atmosphere models directly, thus circumventing the need for both laborious calculations of the energies via HF methods and simultaneously avoiding the need for spline interpolations of tabulated data of binding energies. The maximum errors in the fits are also provided in Table~\ref{tab:Table2}.

\begin{figure}[h]
\begin{center}
\includegraphics[width=3.5in, scale=1.0]{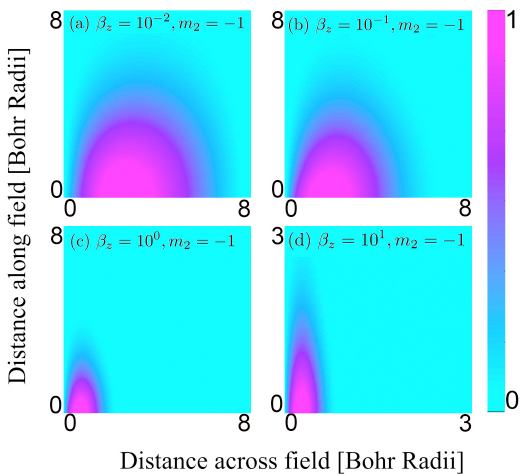}
\end{center}
\caption{Plot of the real part of the wave function, i.e. $\psi(\rho,z)$, of one of the electrons comprising the $M=-1$ state of the helium atom with quantum numbers $m_{2}=-1, s_{z}=-1/2, \pi=+1$ in magnetic fields of strength corresponding to (a) $\beta_{Z}=10^{-2}$, (b) $\beta_{Z}=10^{-1}$, (c) $\beta_{Z}=1$ and (d) $\beta_{Z}=10$. In figure (d) the relevant portion has been enlarged for better visual inspection.}
\label{fig:figure6}
\end{figure}

The calculated data for the states \begin{math} M=-1, -2\end{math} and \begin{math} 0\end{math} with \begin{math} S_{z}=-1\end{math} are compared with the calculated data of the researchers in Refs.~\cite{Ruder94, Jones1996, Jones1999, Ivanov1994, Schmelcher2000, Schmelcher2001, Schmelcher1999} in Table~\ref{tab:Table4} in the appendix. It can be seen upon inspecting Table~\ref{tab:Table4}, that the current findings are consistent with previous work. In particular they are seen here to be improvements upon the estimates of Ruder et al \cite{Ruder94} over the entire range of \begin{math} \beta_{Z}\end{math} considered in this study. Jones et al on the other hand, had employed a Monte-Carlo approach \cite{Jones1996} for solving the HF equations. They assumed a large number of basis functions with variable parameters that could be fine tuned within the framework of Monte-Carlo simulations to arrive at upper bounds for the energies. Though this method is effective, it is computationally demanding and it restricts the wave functions of the electrons to be expressed using a finite number of basis functions. The method described in the current study does not impose such a condition and thus, the wave functions that are determined are, in effect, superpositions of a very large number of such basis functions and arise naturally from the solution itself. The eigenvalues were seen to drop below those of Jones et al's estimates beyond $\beta_{Z} \approx 20$. It is to be noted that this effect was due to insufficient computer memory to carry out calculations with the desired mesh refinement at high magnetic field strengths, where the electrons are tightly bound to the nucleus and their wave functions shrink closer to the nucleus; see Figure~\ref{fig:figure6}. With sufficient computer memory, it is expected that the results presented here can be extended to higher magnetic field strengths.

In Table~\ref{tab:Table4}, it can be seen that the estimates of the binding energy of state of helium corresponding to the configuration $M=-1$, are improvements upon the estimates of Ruder et al \cite{Ruder94}. The range of improvement is between 0.01\% at $\beta_{Z}=10^{-2}$, to about 9.5\% at $\beta_{Z}=1$. These eigenvalues can be seen to be better estimates than those of Jones et al \cite{Jones1996, Jones1999} over the entire range \begin{math} 10^{-2} \leq \beta_{Z} \leq 10^{1} \end{math} , by about 0.5\% maximum and about 0.1\% minimum. With regard to the data obtained by the authors in Ref.~\cite{Schmelcher2000}, it can be seen that the current study does better towards the middle of the range in $\beta_{Z}$ by a maximum of about 0.25\%, while the energies of Ref.~\cite{Schmelcher2000} are seen to be more bound at lower field strengths by a maximum of about 0.12\%. It is to be mentioned that the current work achieves the desired level of accuracy with only a single configuration calculation, without the aid of a basis of functions and without approximations for evaluating the electron potentials.

Additionally, it can be seen in Table~\ref{tab:Table4} that estimates of the binding energies of the first and second excited states of the helium atom in strong magnetic fields, are again consistent with the findings of other researchers \cite{Ruder94, Jones1996, Jones1999, Schmelcher2001, Schmelcher1999}. For the state $M=-2$, the range of improvement was between 10.7\% maximum to about 0.2\% minimum, over the entire range \begin{math} 10^{-2} \leq \beta_{Z} \leq 10 \end{math}, relative to the eigenvalues obtained by the the researchers in Ref.~\cite{Ruder94}. Additionally, the range of improvement relative to the eigenvalues obtained by Jones et al \cite{Jones1996, Jones1999}, can be seen to be 0.6\% maximum to 0.1\% minimum. With regard to the eigenvalues obtained by the authors in Ref.~\cite{Schmelcher2001}, the extent of improvement was between 0.1\% minimum to 0.5\% maximum. Towards higher magnetic field strengths it was observed that the improvements in the estimates tended to drop; this is due to the fact that at such high magnetic field strengths, for obtaining more accurate estimates of the energies, greater computer memory was required to accommodate for finer meshes and this was not possible within the framework of the current project. Similarly, upon examining the data for the third state of helium, $M=0$, it can be seen that the improvements with regard to the data of Ref.~\cite{Ruder94}, was in the range 0.1\% minimum to 17.5\% maximum. With respect to the eigenvalues obtained by Jones et al, the range of improvements of the current data can be seen to be 0.1\% minimum to 0.5\% maximum. Additionally, with regard to the eigenvalues obtained by the authors in Ref.~\cite{Schmelcher1999}, the range of improvements of the current data can be seen to be 0.03\% minimum to 0.35\% maximum. It is to be mentioned that the estimates of the current work straight-forward single configuration calculation are consistent with results from full configuration-interaction studies \cite{Schmelcher1999, Schmelcher2000, Schmelcher2001}, as well as with Monte-Carlo simulations \cite{Jones1996, Jones1999}. Additionally, for the purposes of atmosphere models of neutron stars and white dwarf stars, the models provided in Table~\ref{tab:Table2} could be employed directly yielding accurate eigenvalues in the intermediate range of magnetic field strengths.  
 
Finally, the wave functions for one of the electrons in the state
$M=-1$,$\pi_{Z}=+1$ are plotted in Figure~\ref{fig:figure6}, for four
different values of the magnetic field strength $\beta_{Z}$. For
illustrating the dramatic change in the structure of the atom with
increasing magnetic field strength, we chose to show the electron in
the state with quantum numbers $m=-1, s_{z}=-1/2, \pi_{z}=+1$, which
is equivalent to the 2p$_{-1}$ orbital, in the low field limit. The length
units are Bohr radii of the helium atom,
i.e. \begin{math}a_{B}/Z\end{math} and the horizontal axis represents
the direction perpendicular to the magnetic field, while the
vertical axis represents the direction parallel to the magnetic field, i.e. the $z-$ direction in three dimensional cylindrical co-ordinates. It can be seen upon comparing Figure~\ref{fig:figure6}(d) with Figures~\ref{fig:figure6}(a-c), that the electron wave function shrinks considerably with increasing magnetic field strength and consequently the binding energy increases.

\begin{table}[t]
  \caption{Coefficients of the different rational
    functions for fitting the three states of helium discussed.  The
    maximum fractional error of the eigenvalue relative to the fit from
    $\beta_Z=10^{-2}$ to $\beta_Z=10^1$ was $2 \times 10^{-3}$ in all cases.}

\begin{tabular}{clcl}
\hline
State & \multicolumn{1}{c}{Coefficients} &
State & \multicolumn{1}{c}{Coefficients} \\
\hline
$\begin{array}{cr}
M=-1 \\
1{\rm s}_0;2{\rm p}_{-1} 
\end{array}$ & 
$\begin{array}{cr}
  a_0=&       0.37102955\\
  a_1=&        2.5169555\\
  a_2=&        1.2836647\\
  a_3=&       0.20177864\\
  b_0=&       0.34352813
\end{array}$ &
$\begin{array}{c}
M=-2 \\
1{\rm s}_0;3{\rm d}_{-2} 
\end{array}$ & 
$\begin{array}{cr}
  a_0=&    0.34851215\\
  a_1=&      2.270714\\ 
  a_2=&     1.1867799\\ 
  a_3=&    0.18806319\\
  b_0=&    0.33133095
\end{array}$ \\
\hline
$\begin{array}{c}
M=0 \\
1{\rm s}_0;2{\rm p}_{0} 
\end{array}$ & 
$\begin{array}{cr}
  a_0=&    0.39592729\\
  a_1=&     2.1858926\\
  a_2=&    0.92947212\\
  a_3=&    0.14319581\\
  b_0=&    0.36821346
\end{array}$ & & \\
\hline
\end{tabular}
\label{tab:Table2}
\end{table}

\section{\label{sec:conclusions} Conclusions }

The work described here was motivated by the need to have accurately determined values for the upper bounds of the energy levels of atoms in strong magnetic fields. As was discussed earlier, this need has arisen due to the presence of strong magnetic fields in neutron stars and white dwarf stars. The most commonly present atoms in the atmospheres of these compact objects, hydrogen and helium, were studied here with the intention of obtaining accurate estimates of the energy levels of the first few low lying states, in strong magnetic fields. We described a method adopting a physically motivated approach, governed by the inherent symmetries of the problem. We simultaneously circumvented the need for adopting a definite basis of functions to describe the wave functions of the electrons, in either of the directions, parallel and perpendicular to the magnetic field. The approach is unrestricted with regard to the wave function; it has the distinct advantage over methods that require a basis of functions to describe the wave functions, because in numerical solutions one can only have a finite number of such functions. The wave functions determined in the present study came about naturally from the symmetries of the problem and are thus, in effect, superpositions of a large number of basis functions.

Such an approach resulted in elliptic partial differential equations for the electrons, that were subsequently solved using finite element techniques. It is to be noted that the computational method adopted for determining the direct and exchange interactions between the electrons in the helium atom is also exact, in the sense that it does not rely upon any \emph{ab initio} assumptions to approximate the integrals. These interaction potentials are solved in a natural manner by solving the elliptic partial differential equations, Eqs.~(\ref{eq:14}) and~(\ref{eq:20}). The eigenvalues found in the range of the magnetic field strength parameter \begin{math}10^{-2} \leq \beta, \beta_{Z} \leq 10\end{math} considered in this study were seen to be consistent with previous findings \cite{Ruder94, Jones1996, Jones1999, Ivanov1994, Schmelcher1999, Schmelcher2000, Schmelcher2001}. Rational functions were also used to find sufficiently accurate interpolating functions for the binding energies of various states of both the hydrogen and helium atoms, in the range of magnetic fields considered herein. These were seen to be accurate to (an average for all six fits) within 0.5\%. Potentially, such interpolating functions could be used in atmosphere models of neutron stars and white dwarf stars, thus obviating the need for involved and laborious calculations of the same.

Thus, the current work describes an unrestricted and computationally less intensive method for calculating the energy levels of atoms in strong magnetic fields. There are in essence three directions in which the current work could be extended. First, the current work can be readily extended to higher magnetic field strengths by using adaptive mesh refinement to incorporate the fact that the electrons become increasingly bound. Simultaneously, the calculations and the software developed as a part of this study are readily extendable to systems with more than two electrons. In this regard, it is to be noted that there is only limited work available in the literature for atoms with more than two electrons. At this juncture, the reader is referred to work carried out by Schmelcher and co-workers over recent years on multi-electron atoms in magnetic fields, for accurate data for low-lying states of these systems \cite{Schmelcher_lithium97, Schmelcher_lithium2004, Schmelcher_beryllium2001, Schmelcher_beryllium2004, Schmelcher_boron2001, Schmelcher_carbon99, Schmelcher_sodium2003}. Finally, the procedures implemented herein can also be extended towards a multi-configuration framework \cite{CFF1997}. In essence, the calculations employed herein are for a single configuration of electrons and a multi-configuration approach is likely to improve the results already obtained here.

\begin{acknowledgments}
This research was supported by funding from NSERC.  The calculations
were performed on computing infrastructure purchased with funds from
the Canadian Foundation for Innovation and the British Columbia
Knowledge Development Fund.
\end{acknowledgments}


\bibliography{Paper}

\newpage

\begin{widetext}

\appendix
\section{\label{sec:binding} Tables of Binding Energies }

\begin{table}[h]
\centering
\caption{Binding energies of the three most tightly bound states of hydrogen. Energies are in units of Rydbergs. The states are labelled according to their counterparts in the low-field limit. Results from Ref.~\cite{Ruder94} are also provided for comparison. The number in the parentheses is the absolute error at the fifth decimal place; this is determined as the difference between the computed eigenvalues for the most finely refined mesh size employed and the extrapolated result for the mesh size tending to zero.}
\begin{tabular}{c@{\hspace{5mm}}lc@{\hspace{5mm}}lc@{\hspace{5mm}}ll}
\hline
\hline
$\beta$ & Present Study & Ref. \cite{Ruder94} & Present Study & Ref. \cite{Ruder94} & Present Study & Ref. \cite{Ruder94}\\
& 1s$_{0}$ & 1s$_{0}$ & 2p$_{-1}$ & 2p$_{-1}$ & 3d$_{-2}$ & 3d$_{-2}$\\
\hline
1 $\times$ 10$^{-2}$ & 1.0198(1) &  1.0198 & 0.2876(1) &  0.2876 & 0.1614(0) & 0.1614\\
2 $\times$ 10$^{-2}$ & 1.0392(1) &  1.0392 & 0.3209(1) &  0.3209 & 0.1982(0) & 0.1982\\
5 $\times$ 10$^{-2}$ & 1.0951(1) &  1.0951 & 0.4017(2) &  0.4017 & 0.2757(0) & 0.2757\\
7 $\times$ 10$^{-2}$ & 1.1304(1) &  1.1304 & 0.4452(2) &  0.4452 & 0.3144(1) &	0.3144\\
1 $\times$ 10$^{-1}$ & 1.1808(1) &  1.1808 & 0.5010(3) &  0.5011 & 0.3626(1) & 0.3626\\
2 $\times$ 10$^{-1}$ & 1.3293(1) &  1.3292 & 0.6427(5) &  0.6427 & 0.4820(2) &	0.4820\\
5 $\times$ 10$^{-1}$ & 1.6624(1) &  1.6623 & 0.9132(20) &  0.9132 & 0.7061(9) &	0.7061\\
7 $\times$ 10$^{-1}$ & 1.8324(0) &  1.8323 & 1.0420(34) &  1.0420 & 0.8125(18) &	0.8125\\
1 & 2.0445(2) &  2.0444 & 1.1992(60)	& 1.1992 & 0.9423(36) & 0.9423\\
2 & 2.5616(3) &  2.5616 & 1.5756(37)	& 1.5757 & 1.2540(21) & 1.2540\\
5 & 3.4956(5) &  3.4956 & 2.2507(188) & 2.2508 & 1.8164(122) & 1.8164\\
7 & 3.9225(6) &  3.9224 & 2.5604(353)	& 2.5605 & 2.0758(230) & 2.0758\\
10 & 4.4308(14) & 4.4308 & 2.9303(630) & 2.9310 & 2.3872(451) & 2.3873\\
\hline
\hline
\end{tabular}
\label{tab:Table3}
\end{table}

\newpage

\begin{table}[h]
\centering
\caption{Binding energies of the three states of helium $M=-1, S_{z}=-1, \pi_{z}=+1$; $M=-2, S_{z}=-1, \pi_{z}=+1$ and $M=0, S_{z}=-1, \pi_{z}=-1$. Energies are in units of Rydberg energies in the Coulomb potential of nuclear charge $Ze$, where $Z=2$ for helium. Accurate data from other work is also provided for comparison.}

\begin{tabular}{c@{\hspace{5mm}}c@{\hspace{5mm}}l@{\hspace{5mm}}c@{\hspace{5mm}}c@{\hspace{5mm}}c@{\hspace{5mm}}c@{\hspace{5mm}}l}
\hline
\hline
State & $\beta_{Z}$ & Present Study & Ref.~\cite{Ruder94} & Ref.~\cite{Jones1996} & Ref.~\cite{Jones1999} & Ref.~\cite{Ivanov1994} & Ref.~\cite{Schmelcher2000} \\
\hline
& 1 $\times$ 10$^{-2}$ & 1.1183(1) & 1.1182 & 1.1183 & 1.1183 & 1.1183 & 1.1193\\
& 2 $\times$ 10$^{-2}$ & 1.1612(1)	& 1.1609 &	&	&  & 1.1626\\
& 5 $\times$ 10$^{-2}$ & 1.2691(3)	& 1.2658 & 1.2683 & 1.2683 & & 1.2704\\
& 7 $\times$ 10$^{-2}$ & 1.3319(5) & 1.3258 & 1.3303 &	 &	& \\
& 1 $\times$ 10$^{-1}$ & 1.4189(0)	& 1.4069 & 1.4150 & 1.4151 & 1.4151 & 1.4178\\
$M=-1$ & 2 $\times$ 10$^{-1}$ & 1.6585(3) & 1.6270 & 1.6509 & 1.6508 & 1.6511 & 1.6544 \\
$1s_{0}2p_{-1}$ & 5 $\times$ 10$^{-1}$ & 2.1550(22) & 2.0508 & 2.1475 & 2.1490 & 2.1492 &\\
& 7 $\times$ 10$^{-1}$ & 2.4029(13) & 2.2329 & 2.3927 & 2.3955 &  2.3960 &\\
& 1 & 2.7026(32) & 2.4675 & 2.6897 & 2.7000 & 2.7003 &\\
& 2 & 3.4384(75) & 3.2394	 & 	         & 3.4333 & &\\
& 5 & 4.7502(314) & 4.5899 &	         & 4.7441 & &\\
& 7 & 5.3474(556) &	          &	         & 5.3408 &     &         \\
& 10 & 6.0543(791) & 5.9206 & 	         & 6.0506 &  6.0507 & \\
\hline
State & $\beta_{Z}$ & Present Study & Ref.~\cite{Ruder94} & Ref.~\cite{Jones1996} & Ref.~\cite{Jones1999} & Ref.~\cite{Schmelcher2001}\\
\hline
& 1 $\times$ 10$^{-2}$ & 1.0852(518)	& 1.0828 & 1.0832 & 1.0830 & 1.0833\\
& 2 $\times$ 10$^{-2}$ & 1.1234(228) & 1.1207 &  & & 1.1224\\
& 5 $\times$ 10$^{-2}$ & 1.2175(154) & 1.2097 & 1.2155 & 1.2160 & 1.2167\\
& 7 $\times$ 10$^{-2}$ & 1.2732(165) & 1.2596 & 1.2690 & & \\
& 1 $\times$ 10$^{-1}$ & 1.3510(134)	&1.3266 & 1.3412  & 1.3436& 1.3450\\
$M=-2$ & 2 $\times$ 10$^{-1}$ & 1.5598(75) & 1.5073 & 1.5401 & 1.5503 & 1.5525\\
$1s_{0}3d_{-2}$ & 5 $\times$ 10$^{-1}$ & 2.0009(740) & 1.8508 & 1.9806 & 1.9945 & \\
& 7 $\times$ 10$^{-1}$ & 2.2246(937)	&1.9935 & 2.1800 & 2.2171  & \\
& 1 & 2.4981(845) & 2.2572 & 2.4362 & 2.4933 & \\
& 2 & 3.1685(889) & 2.9683 &  & 3.1634 & \\
& 5 & 4.3740(1035) & 4.2161 &	& 4.3693 & \\
& 7 & 4.9268(1647) &  & & 4.9203 & \\
& 10 & 5.5851(3733) & 5.4492 &  &	 5.5770 & \\
\hline
State & $\beta_{Z}$ & Present Study & Ref.~\cite{Ruder94} & Ref.~\cite{Jones1996} & Ref.~\cite{Jones1999} & Ref.~\cite{Schmelcher1999} \\
\hline
& 1 $\times$ 10$^{-2}$ & 1.1029(9) & 1.1016 & 1.1018 & 1.1016 & 1.1026 \\
& 2 $\times$ 10$^{-2}$ & 1.1336(41)	& 1.1314 &	&	& 1.1333 \\
& 5 $\times$ 10$^{-2}$ & 1.2135(80)	& 1.2028 & 1.2098 & 1.2099 & 1.2112 \\
& 7 $\times$ 10$^{-2}$ & 1.2573(17) & 1.2421 & 1.2551 &	 &	\\
& 1 $\times$ 10$^{-1}$ & 1.3231(15)	& 1.2933 & 1.3174 & 1.3174 & 	1.3191\\
$M=0$ & 2 $\times$ 10$^{-1}$ & 1.4988(25) & 1.4197 & 1.4914 & 1.4914 & 1.4936\\
$1s_{0}2p_{0}$ & 5 $\times$ 10$^{-1}$ & 1.8647(11) & 1.5864 & 1.8571 & 1.8593 & \\
& 7 $\times$ 10$^{-1}$ & 2.0461(13) & 1.7833 & 2.0349 & 2.0404 &  \\
& 1 & 2.2685(16) & 2.0304 & 2.2431 & 2.2641 & \\
& 2 & 2.8060(12) & 2.6037	 & 	         & 2.7989 & \\
& 5 & 3.7522(1098) & 3.5916 &	         & 3.7466 & \\
& 7 & 4.1817(323) &	          &	         & 4.1762 &              \\
& 10 & 4.6920(495) & 4.5560 & 	         & 4.6862 &               \\
\hline
\hline
\end{tabular}
\label{tab:Table4}
\end{table}
\end{widetext}


\end{document}